\documentclass{llncs}

\usepackage{float}
\usepackage{url}
\usepackage{amssymb}

\usepackage{url}

\usepackage{graphicx}              
\usepackage[latin2]{inputenc}
\usepackage{mathtools}
\usepackage{ifthen}
\usepackage{array}
\usepackage{tikz}
\usetikzlibrary{decorations.pathreplacing,calc}
\usepackage{verbatim}
\usepackage{arydshln}
\usepackage{float}
\usepackage{array}

\usepackage{times}

\newcommand{\draw}{\leftarrow}

\newcommand{\inp}{\textbf{in:}}

\newcommand{\maxBB}{\mathsf{max_{BB}}}
\newcommand{\BODY}[1]{[#1]}
\newcommand{\body}{\mathit{body}}

\newcommand{\sR}{\mathsf{R}}
\newcommand{\sA}{\mathsf{A}}
\newcommand{\sB}{\mathsf{B}}
\newcommand{\sC}{\mathsf{C}}

\newcommand{\graphC}[1]{\begin{center}\graph{#1}\end{center}}

\newcommand{\out}[1]{}

\newcommand{\up}{\vspace{-0.1truecm}}

\tikzset{>=latex}

\newcommand{\graph}[1]
{
  {
  
    \footnotesize
    \centering
    \begin{tikzpicture}[
      nodes={
	inner sep=-0.5\pgflinewidth, outer sep=-0.5\pgflinewidth,
	inner xsep=-3.5\tabcolsep, 
	outer xsep=0.5\pgflinewidth, 
	align=left
      }]
      #1
    \end{tikzpicture}
    
  }
}

\newcommand{\desc}[2][0.95]
{
  \centering
  \fbox{ 
    \footnotesize
    \begin{minipage}[c]{#1\linewidth}
      #2
    \end{minipage}
  }
}

\newcommand{\transactionAux}[7]{
  \begin{tabular}{|m{#2 cm}|}
    \hline
    \multicolumn{1}{|c|}{\textbf{#3}} \\
    \hline
    #4 \\
    \hline
    \textbf{out-script(#5):} #6 \\
    \textbf{val:} #7
    \ifthenelse{\not \equal{#1}{0}}
      {\\ \hline
	\textbf{tlock:} #1 \\
	\hline}
      {\\ \hline}
  \end{tabular}
}

\newcommand{\DtransactionAux}[7]{
  \begin{tabular}{:m{#2 cm}:}
    \hdashline
    \multicolumn{1}{:c:}{\textbf{#3}} \\
    \hdashline
    #4 \\
    \hdashline
    \textbf{out-script(#5):} #6 \\
    \textbf{val:} #7
    \ifthenelse{\not \equal{#1}{0}}
      {\\ \hdashline
	\textbf{tlock:} #1 \\
	\hdashline}
      {\\ \hdashline}
  \end{tabular}
}

\newcommand{\Drect}[4]{
  \node[rectangle] (#3) at (#1) 
  {\begin{tabular}{|c|} \hline \\ \hspace{#2 cm} #4 \hspace{#2 cm} \\ \\ \hline \end{tabular}};
}

\newcommand{\DtransTwoInTwoOut}[9]{
  \node[rectangle] (#3) at (#1)
  {\DtransactionAuxTwoInTwoOut{#2}{#4}{#5}{#6}{#7}{#8}{#9}};
}

\newcommand{\DtransactionAuxTwoInTwoOut}[7]{
  \begin{tabular}{|m{#1 cm}|}
    \hline
    \multicolumn{1}{|c|}{\textbf{#2}} \\
    \hline
    \begin{tabular}{m{0.48\linewidth} | m{0.47\linewidth}}
        #3
    \end{tabular}
    \\
    \hline
    \begin{tabular}{m{0.48\linewidth} | m{0.47\linewidth}}
      \textbf{out-script$_1$(#4):} #5 &
      \textbf{out-script$_2$(#4):} #6 \\
      \textbf{val$_1$:} #7 & \textbf{val$_2$:} #7
    \end{tabular}
    \\
    \hline
  \end{tabular}
}

\newcommand{\transaction}[7][0]{
  \transactionAux{#1}{#2}{#3}{#4}{#5}{#6}{#7}
}

\newcommand{\Dtransaction}[7][0]{
  \DtransactionAux{#1}{#2}{#3}{\textbf{in-script:} #4}{#5}{#6}{#7}
}

\newcommand{\trans}[9][0]{
  \node[rectangle] (#4) at (#2)
  {\transaction[#1]{#3}{#5}{\textbf{in-script:} #6}{#7}{#8}{#9}};
}

\newcommand{\Dtrans}[9][0]{
  \node[rectangle] (#4) at (#2)
  {\Dtransaction[#1]{#3}{#5}{#6}{#7}{#8}{#9}};
}

\makeatletter
\newcommand{\thickhline}{%
    \noalign {\ifnum 0=`}\fi \hrule height 1pt
    \futurelet \reserved@a \@xhline
}
\newcolumntype{"}{@{\hskip\tabcolsep\vrule width 1pt\hskip\tabcolsep}}
\makeatother

\newcommand{\BTC}{%
  \leavevmode
  \vtop{\offinterlineskip 
    \setbox0=\hbox{B}%
    \setbox2=\hbox to\wd0{\hfil\hskip-.03em
    \vrule height .3ex width .15ex\hskip .08em
    \vrule height .3ex width .15ex\hfil}
    \vbox{\copy2\box0}\box2}}

\newcommand{\sig}{\mathsf{sig}}
\newcommand{\ver}{\mathsf{ver}}

\newcommand{\Commit}{\mathit{Commit}}
\newcommand{\Reveal}{\mathit{Open}}
\newcommand{\FuseAR}{\mathit{Fuse}}

\newcommand{\CommitSCS}{\mathit{Commit}}
\newcommand{\RevealSCS}{\mathit{Open}}
\newcommand{\FuseSCS}{\mathit{Fuse}}

\newcommand{\Open}{\mathit{Open}}
\newcommand{\Fuse}{\mathit{Fuse}}

\newcommand{\PutMoney}{\mathit{Deposit}}

\newcommand{\FC}{{\mathsf{FairComputation}}}

\newcommand{\CS}{\mathsf{CS}}
\newcommand{\CSCom}{{\mathsf{CS.Commit}}}
\newcommand{\AR}{{\mathsf{CS}}}
\newcommand{\ARCom}{{\mathsf{CS.Commit}}}
\newcommand{\AROp}{{\mathsf{CS.Open}}}
\newcommand{\SCS}{{\mathsf{SCS}}}

\newcommand{\NSCS}{{\mathsf{NewSCS}}}
\newcommand{\NSCSCom}{{\mathsf{NewSCS.Commit}}}
\newcommand{\NSCSOp}{{\mathsf{NewSCS.Open}}}

\begin{document}

\title{How to deal with malleability of BitCoin transactions}
\author{Marcin~Andrychowicz\thanks{marcin.andrychowicz@crypto.edu.pl},
Stefan~Dziembowski\thanks{stefan.dziembowski@crypto.edu.pl},
Daniel~Malinowski\thanks{daniel.malinowski@crypto.edu.pl} and
{\L}ukasz~Mazurek\thanks{lukasz.mazurek@crypto.edu.pl}}
\institute{University of Warsaw}

\maketitle

\begin{abstract}
 BitCoin transactions are malleable in a sense that given a transaction 
an adversary can easily construct an equivalent transaction which has a different hash.
This can pose a serious problem in some BitCoin distributed contracts in which 
changing a transaction's hash may result in the protocol disruption and a financial loss.
The problem mostly concerns protocols, which use a ''refund'' transaction to withdraw a deposit in a case of
the protocol interruption.
In this short note, we show a general technique for creating malleability-resilient ``refund'' transactions,
which does not require any modification of the BitCoin protocol.

Applying our technique to our previous paper ``Fair Two-Party Computations via the BitCoin Deposits'' (Cryptology ePrint Archive, 2013)
allows to achieve fairness in any Two-Party Computation using the BitCoin protocol in its current version.

\end{abstract}

\section{Malleability of BitCoin transactions}
We assume that the reader is familiar with the BitCoin
protocol and in particular with non-standard transaction scripts (used e.g. in 
so-called \emph{distributed contracts}).
For general description of BitCoin, see e.g. \cite{nakamoto,ADMM13b} or BitCoin wiki
page \url{http://en.bitcoin.it/}.
For the description of non-standard transaction scripts, see \cite{ADMM13,ADMM13b}
or \emph{Contracts} page \url{http://en.bitcoin.it/wiki/Contracts}.

BitCoin transactions are malleable\footnote{See \url{http://en.bitcoin.it/wiki/Transaction_Malleability}.}
in a sense that given
a transaction $T$ it is easy to create a functionally identical transaction $T'$
($T$ and $T'$ differs only in the input scripts)
which has a different hash\footnote{This can be done e.g. by adding \emph{push} and
\emph{pop} commands to the input script}.
This gives an adversary an opportunity to slightly change the transaction sent
by a user before it is included in the blockchain.
It strongly affects the \emph{distributed contracts} which use the hashes of
the transactions before broadcasting them.

The source of the malleability is the fact that in the current version of the BitCoin protocol,
each transaction contains a hash of the \emph{whole} transaction it spends,
while the signatures are taken over 
the simplified version of the transaction (excluding
the input scripts).

The most common scenario in which the malleability of transactions is a problem
is the following. Suppose that there is a transaction $\PutMoney$, which
should be redeemed by a transaction $\Fuse$\footnote{Transactions of this kind are sometimes called \emph{refund} transactions.} with
time-lock $t$, but
for some reason $\Fuse$ has to be created and signed before $\PutMoney$
is broadcast.\footnote{See e.g. examples 1, 5 and 7 
on \url{http://en.bitcoin.it/wiki/Contracts}.}.
In the above scenario a problem arises if the $\PutMoney$ transaction is maliciously changed
and its version included in the blockchain has a different hash than expected, what invalidates
the transaction $\Fuse$.

In our recent paper \cite{ADMM13b}
we proposed a modification of BitCoin which eliminates the malleability problem.
The idea of this modification was to identify the transactions by the hashes of their \emph{simplified}
versions (excluding the input scripts).
With this modification one can of course still modify the input script of the transaction,
but the modified transaction would have the same hash.
We used this improvement of BitCoin to guarantee
the correctness of the $\Fuse$ transactions,
which had to be sign before broadcasting its input transaction.
In this short note we present another approach to achieving the correctness of
$\Fuse$ transactions which does not need any modification of the BitCoin protocol.


\section{New technique}


Our technique uses a \emph{BitCoin-based timed commitment scheme} introduced in \cite{ADMM13}.
We briefly describe this commitment scheme in Sec.~\ref{sec:CS}.
Later in Sec.~\ref{sec:trick} we show how to construct $\Fuse$ transactions, which are resistant to malleability
and in Sec.~\ref{sec:SCS} we apply it to $\SCS$ protocol
from \cite{ADMM13b}, what leads to a general fair Two-Party Computation protocol, which
is secure in the current version of the BitCoin protocol (in particular, even if transactions are malleable).
In Sec.~\ref{sec:other} we list other protocols, which can be made resistant to malleability using our technique.

\subsection{BitCoin-based timed commitment scheme}\label{sec:CS}

\begin{figure*}[h]
  \graphC{
    \input{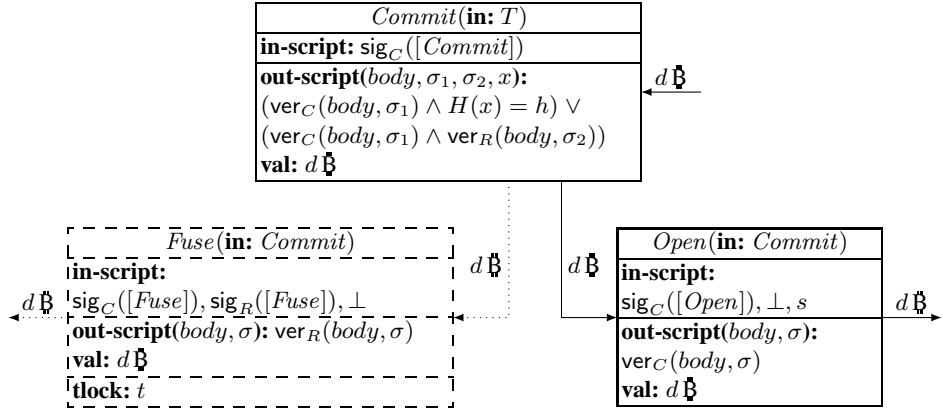}
  }
  
  \vspace{.3truecm}
  
  \desc[0.9]{
    \input{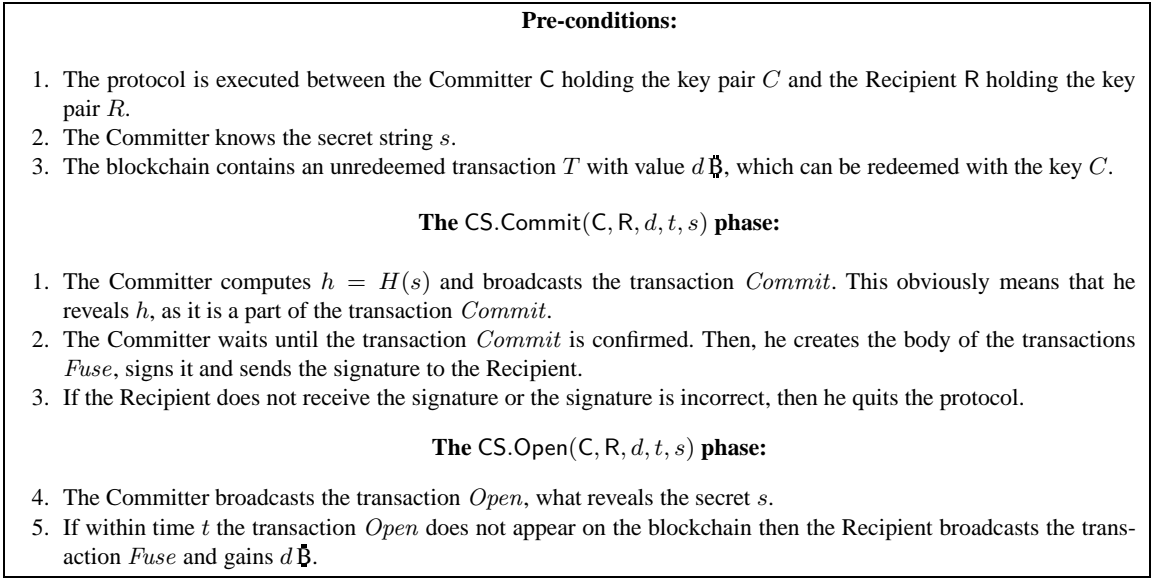}
  }
\caption{The $\AR$ protocol. The scripts' arguments, which are omitted are denoted by $\bot$.}
\label{fig:AR}
\end{figure*}

In this section we briefly describe a timed commitment scheme
from \cite{ADMM13}, which will be denoted $\AR$.
The protocol $\AR$ is executed between the Committer (denoted $\sC$)
and the Recipient (denoted $\sR$).
During the commitment phase the Committer commits himself to some string $s$ by
revealing its hash $h = H(s)$.
Moreover the parties agree on a moment of time $t$ until which the Committer should open the commitment, i.e.
reveal the secret value $s$.
The protocol is constructed in such a way that if the Committer does not open the commitment until time $t$, then
the agreed amount of $d\,\BTC$ is transfered from the Committer to the Recipient.
More precisely, at the beginning of the protocol the Committer makes a deposit of $d\,\BTC$, which
is returned to him if he opens the commitment before time $t$ or taken by the Recipient otherwise.

We follow the notation from \cite{ADMM13,ADMM13b} in which
the transactions are represented as boxes.
The graph of transactions and the full description of the $\CS$ protocol is presented on Fig.~\ref{fig:AR}.
Refer to \cite{ADMM13} for more details.
Notice that even if the transaction $\Commit$ is maliciously changed before being included in the block,
the protocol still succeeds because the transaction $\Fuse$ is created after $\Commit$ is included in the blockchain,
so it always contains the correct hash of $\Commit$.
Therefore, the $\CS$ protocol is resistant to transaction malleability.

The execution of the commitment phase with $\sC$ as the Committer and $\sR$ as the Recipient
will be denoted by $\ARCom(\sC, \sR, d, t, s)$, where $d$ is the size of the deposit and $t$ is the time until which $\sC$
should reveal the secret $s$.

\subsection{$\Fuse$ transactions resistant to malleability}\label{sec:trick}

\begin{figure}[ht]
 \graphC{
   \input{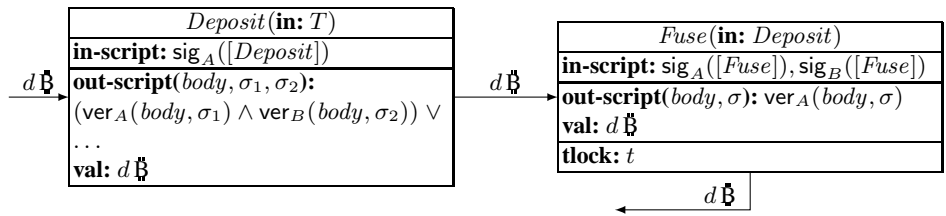}
 }
\up\up
\up\up

 \caption{The typical solution with the $\Fuse$ transaction vulnerable to malleability.}

\label{fig:notbot}
\end{figure}

Suppose that in the execution of some protocol between the parties $\sA$ and $\sB$
there is a transaction $\PutMoney$, which
should be redeemed to an address controlled by $\sA$ at the time $t$
if it is not spent earlier.\footnote{See e.g. examples 1, 5 and 7 
on \url{http://en.bitcoin.it/wiki/Contracts}.}
The typical solution would be to create a transaction
$\Fuse$ with time-lock $t$, which is signed by both parties and redeems $\PutMoney$.
Moreover, $\PutMoney$ has to be claimable using signatures of both parties.
The graph of transactions for this situation is presented on Fig.~\ref{fig:notbot}.

\up\up
\up\up
\up\up
\up\up
\up\up

\begin{figure}[h!]
 \graphC{
   \input{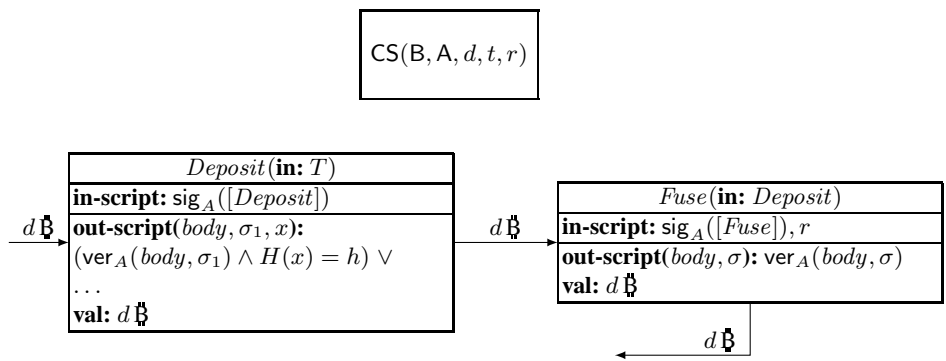}
 }
\caption{The solution with the $\Fuse$ transaction resistant to malleability.
$\AR(\sB, \sA, d, t, r)$ denotes the transactions in the appropriate execution of the $\AR$ protocol.}

\label{fig:notbot2}
\end{figure}

We will now present a technique for creating $\Fuse$ transactions, which are resistant to malleability.
The general idea is to use a timed commitment instead of using a time-lock directly.
More precisely, the transaction $\PutMoney$ should be claimable with a signature of $\sA$
and a random secret $r$, which is known only to $\sB$.
It means that $\sA$ can claim $\PutMoney$ using the $\Fuse$ transaction
as soon as the secret string $r$ is revealed by $\sB$.
In our situation we would like the secret to be revealed at the time $t$.
It can be achieved by executing at the very beginning (before broadcasting the $\PutMoney$ transaction)
the $\ARCom(\sB, \sA, d, t, r)$ protocol.
In this case at the time $t$ either: the $\AR$ commitment was opened, the secret $r$ is known and $\sA$ can broadcast the $\Fuse$ transaction (assuming that $\PutMoney$
was not spent earlier)
or the $\AR$ commitment was not opened and $\sA$ gets $d\,\BTC$ from the $\Fuse$ transaction in the $\AR$ execution.
The graph of transactions is presented on Fig.~\ref{fig:notbot2}.

\up\up
\up\up

The key difference, which makes a new construction resistant to malleability is that the $\Fuse$ transaction does not need
to be signed by $\sB$, so it can be created and signed by $\sA$ \emph{after} the $\PutMoney$ transaction is confirmed
and its hash is known.
A drawback of this construction is that the other party ($\sB$ in our case) also has to make a deposit.




\subsection{Fair Two-Party Computation protocol}\label{sec:SCS}

\begin{figure}[ht]
 \graphC{
    \input{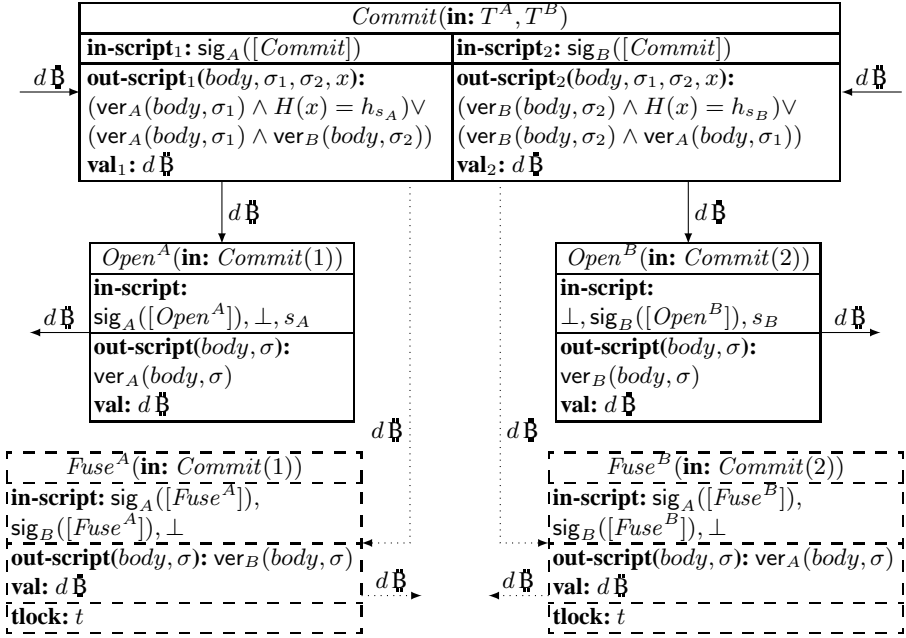}
 }
\caption{The graph of transactions for the original version of the $\SCS$ protocol.}
\label{fig:SCS-orig}
\end{figure}

The \emph{simultaneous BitCoin-based timed commitment scheme} ($\SCS$) described in
\cite{ADMM13b} is an extended version of the $\CS$ protocol in which two parties
simultaneously commit to their secret strings.
The pivotal property of this protocol is that after the commitment phase either:
both parties are committed or none of them is committed (the latter is only possible if one of the parties misbehaved).
The graph of transactions is presented on Fig.~\ref{fig:SCS-orig}.
Refer to \cite{ADMM13b} for more details.

The main application of the $\SCS$ protocol is the $\FC$ protocol from \cite{ADMM13b}, which
is a general fair Two-Party Computation protocol (refer to \cite{ADMM13b} for more details).
However, in contrast to $\CS$, the $\SCS$ protocol is
vulnerable to transaction malleability, because it requires the $\Fuse$ transactions
to be created and signed before broadcasting their input transaction.
Therefore, in \cite{ADMM13b} it was assumed that the BitCoin protocol is modified in such a way, that the transactions
are no longer malleable.

In this section we present a modified version of $\SCS$ protocol called $\NSCS$,
which is resistant to transactions malleability and does not require any change of the BitCoin protocol.
Combining it with the $\FC$ protocol from \cite{ADMM13b} it gives the general fair Two-Party Computation protocol.


\begin{figure}[h]
 \graphC{
    \input{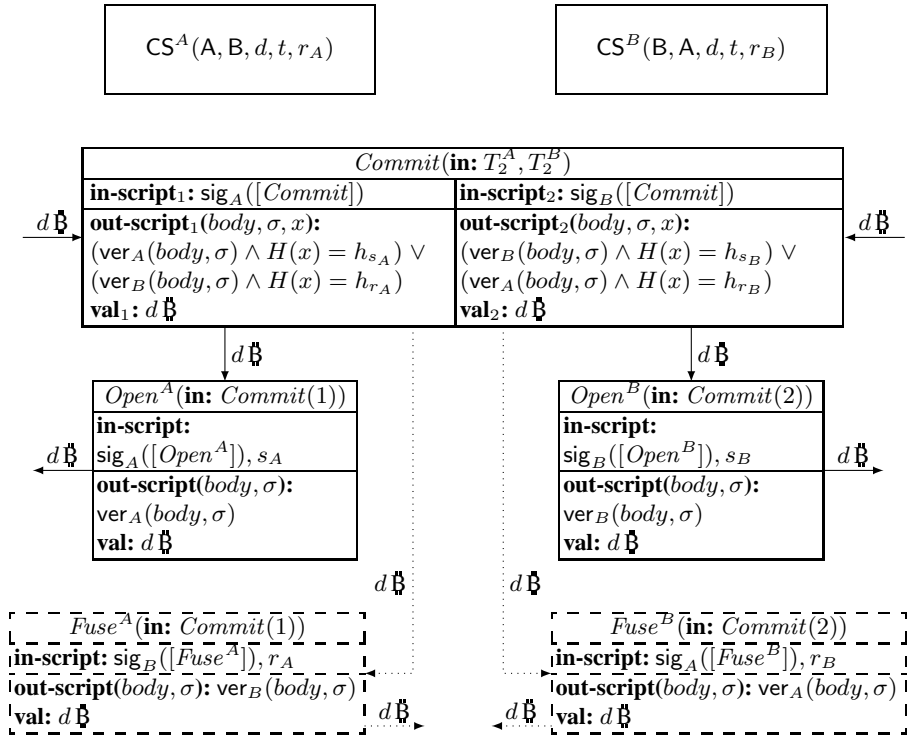}
 }
\caption{The graph of transactions for the $\NSCS$ protocol.
Two boxes labeled with $\AR(\ldots)$ denote the transactions broadcast in the appropriate execution of the $\AR$ protocol.
$h_x$ denotes the value $H(x)$, but it is used in the output scripts to stress
that the value of the hash is directly included in the transaction (instead of value of $x$ and an application of the hash function).}
\label{fig:SCS-graph}
\end{figure}

\begin{figure}[t]
  \desc{
    \input{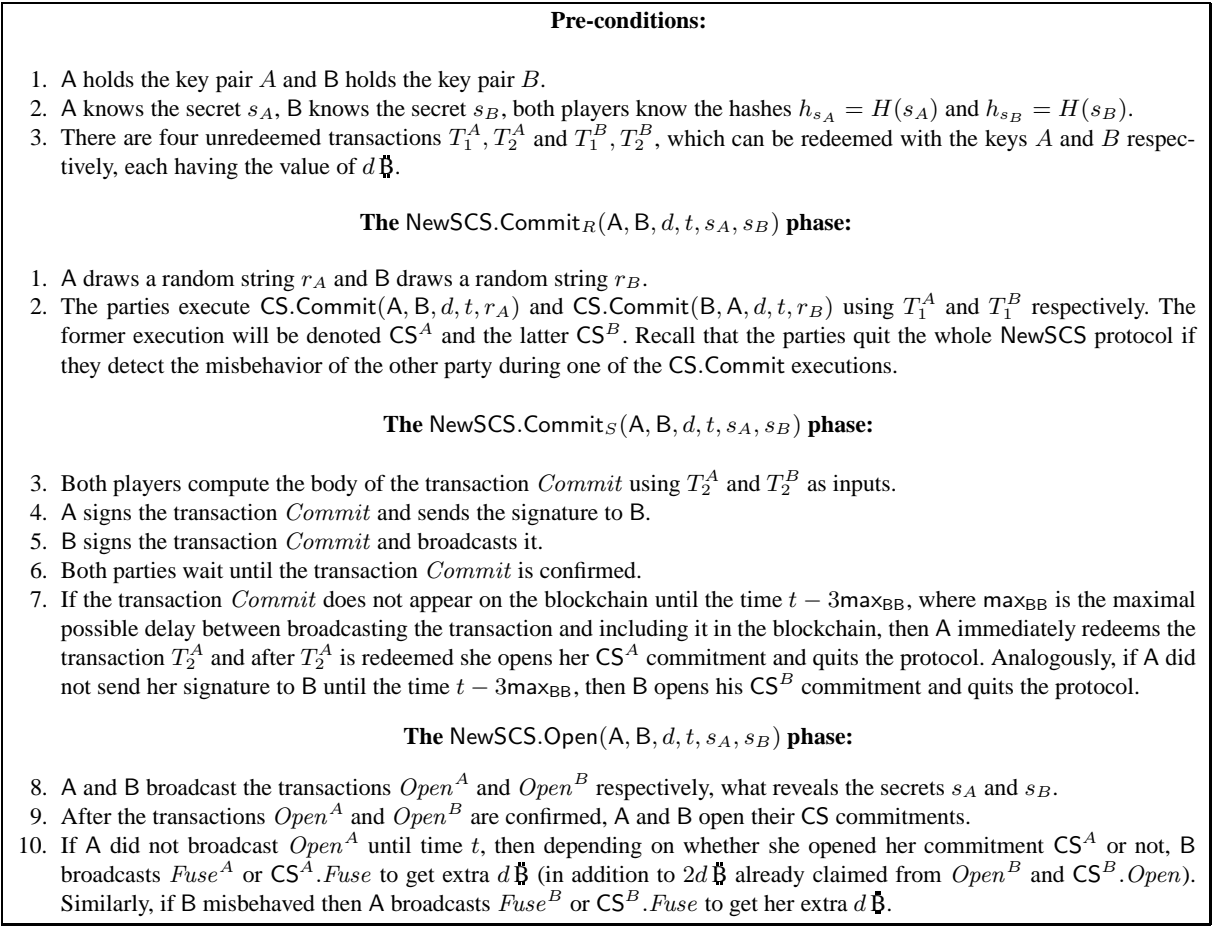}
  }
\caption{The description of the $\NSCS$ protocol.}
\label{fig:SCS-desc}
\end{figure}

The $\NSCS$ protocol is a result of a straightforward application of the technique from Sec.~\ref{sec:trick}
to the $\SCS$ protocol. The graph of the transactions and the full description of the
$\NSCS$ protocol are presented on Fig.~\ref{fig:SCS-graph} and Fig.~\ref{fig:SCS-desc}.



\subsection{Other applications}\label{sec:other}

In this section we list some other protocols, which can be made resistant to malleability
using our technique:
\begin{itemize}
 \item \url{http://en.bitcoin.it/wiki/Contracts}, \emph{Example 1: Providing a deposit}.
 Although, this protocol could be fixed using our technique, the resulting protocol would be rather impractical
 as it would require the server to also make a deposit.
 \item \url{http://en.bitcoin.it/wiki/Contracts}, \emph{Example 5: Trading across chains}.
 \item \url{http://en.bitcoin.it/wiki/Contracts}, \emph{Example 7: Rapidly-adjusted (micro)payments to a pre-de\-ter\-mined party}.
 \item Back and Bentov's lottery protocol from \cite{BB13}.
\end{itemize}

\bibliographystyle{plain}

\bibliography{paper,conf,abbrev_short,crypto}

\end{document}